\documentclass[conference]{IEEEtran}
\usepackage{amsmath,amsfonts}
\usepackage{amssymb}
\usepackage{algorithmic}
\usepackage{algorithm}
\usepackage{array}
\usepackage{color}
\usepackage{booktabs} 
\usepackage[caption=false,font=footnotesize,labelfont=rm,textfont=rm]{subfig}
\usepackage{textcomp}
\usepackage{stfloats}
\usepackage{makecell}
\usepackage{url}
\usepackage{verbatim}
\usepackage{graphicx}
\usepackage{cite}
\newcommand{\figref}[1]{Fig. \ref{#1}}
\newcommand{\tabref}[1]{Table \ref{#1}}

\begin{document}

\title{Energy Efficient Trajectory Control and Resource Allocation in Multi-UAV-assisted MEC via Deep Reinforcement Learning}

\author{\IEEEauthorblockN{Saichao Liu\IEEEauthorrefmark{2}, Geng Sun\IEEEauthorrefmark{2}\IEEEauthorrefmark{3}, Chuang Zhang\IEEEauthorrefmark{2}, 
Xuejie Liu\IEEEauthorrefmark{2},
Jiacheng Wang\IEEEauthorrefmark{3},
Changyuan Zhao\IEEEauthorrefmark{3},
 and Dusit Niyato\IEEEauthorrefmark{3}}

\IEEEauthorblockA{\IEEEauthorrefmark{2}{College of Computer Science and Technology, Jilin University, Changchun 130012, China} \\
\IEEEauthorrefmark{3}{College of Computing and Data Science, Nanyang Technological University, Singapore 639798, Singapore}}
}  

\thanks{This paper was produced by the IEEE Publication Technology Group. They are in Piscataway, NJ.}
\thanks{Manuscript received April 19, 2021; revised August 16, 2021.}

\markboth{Journal of \LaTeX\ Class Files,~Vol.~14, No.~8, August~2021}%
{Shell \MakeLowercase{\textit{et al.}}: A Sample Article Using IEEEtran.cls for IEEE Journals}

\IEEEpubid{0000--0000/00\$00.00~\copyright~2021 IEEE}

\maketitle

\begin{abstract}
Mobile edge computing (MEC) is a promising technique to improve the computational capacity of smart devices (SDs) in Internet of Things (IoT). However, the performance of MEC is restricted due to its fixed location and limited service scope. Hence, we investigate an unmanned aerial vehicle (UAV)-assisted MEC system, where multiple UAVs are dispatched and each UAV can simultaneously provide computing service for multiple SDs. To improve the performance of system, we formulated a UAV-based trajectory control and resource allocation multi-objective optimization problem (TCRAMOP) to simultaneously maximize the offloading number of UAVs and minimize total offloading delay and total energy consumption of UAVs by optimizing the flight paths of UAVs as well as the computing resource allocated to served SDs. Then, consider that the solution of TCRAMOP requires continuous decision-making and the system is dynamic, we propose an enhanced deep reinforcement learning (DRL) algorithm, namely, distributed proximal policy optimization with imitation learning (DPPOIL). This algorithm incorporates the generative adversarial imitation learning technique to improve the policy performance. Simulation results demonstrate the effectiveness of our proposed DPPOIL and prove that the learned strategy of DPPOIL is better compared with other baseline methods. 
\end{abstract}


\section{Introduction}
\par With the rapid development of Internet of Things (IoT) and communication technology, smart devices (SDs) become indispensable in various fields, which can detect objects in autonomous control or be regarded as meteorological sensors to monitor environment \cite{wang2020optimal}. SDs can be deployed to monitor and collect data from relevant areas, thus creating the conditions for emerging intelligent applications. However, these applications are often computing-intensive, which will lead to a huge demand for computing resource. Therefore, SDs face great challenge as their computing resource and battery capacity are limited.
 
\par Mobile edge computing (MEC) is equipped with the substantial computational resource at the edge of networks \cite{mach2017mobile}. Then, SDs can offload computing-intensive applications to the nearby ground base stations (BSs), which can improve the processing efficiency of applications and save the energy consumption of SDs. However, the traditional MEC may not always lead to the satisfying computation offloading performance, which is restricted due to its fixed location and limited communication coverage \cite{zhou2020deep}. Moreover, the BSs may be destroyed due to the military actions or natural disasters, resulting in shortage of computing resource and offloading performance degradation \cite{zhang2020energy}.  

\par Unmanned aerial vehicles (UAVs) achieve the great development that has been widely applied to different domains. With the advantages of low cost, high flexibility and line-of-sight (LoS) connection, UAVs have excellent capability to expand communication coverage. Equipped with MEC server, the UAV can fly around the SDs to collect tasks and perform related offloading service. Nevertheless, UAV is often powered by limited battery, thus constantly adjusting positions and computing tasks of UAV will cause much energy consumption. Therefore, it is significant to optimize the energy consumption of UAV.

\par Traditional methods, such as convex optimization and game theory, have been utilized to optimize the energy consumption of UAV in many communication-related tasks \cite{zeng2019energy} \cite{ruan2018energy}. However, the improvement to MEC performance is in conflict with UAV energy consumption optimization, and multiple optimization variables exist that are non-convex with the optimization objectives. Moreover, the system is dynamic whose state frequently changes. Therefore, conventional methods will face many challenges when solving the communication problem, resulting in struggling to find optimal solutions and incurring high computational costs.

\par Deep reinforcement learning (DRL) has been proven as the efficient optimizer to solve the high-dimensional continuous control problems \cite{fujimoto2018addressing}. Benefiting from deep neural networks (DNNs), DRL can learn the powerful strategy and the agent can perform consecutive decisions without requiring much computational resource. Furthermore, even if the environment changes, the agent can still execute effective decision according to the change in state. Moreover, DRL can be combined with the multi-agent paradigm so as to address the problems requiring coordination among multiple individuals. Therefore, we aim to investigate a UAV-based MEC system based on DRL to achieve intelligent and efficient UAV trajectory control and resource management. The contributions of this work are outlined as follows.
\begin{itemize}
	\item We consider a UAV-assisted MEC system, where multiple UAVs are deployed, and each UAV can simultaneously provide the computing service for multiple SDs. Then, we formulated a UAV-based trajectory control and resource allocation multi-objective optimization problem (TCRAMOP) aimed at jointly maximizing the offloading number of UAVs and minimizing total offloading delay and total energy consumption of UAVs by optimizing the flight paths of UAVs and the allocated computing resource of served SDs. 
	
	\item We propose a DRL algorithm, called distributed proximal policy optimization with imitation learning (DPPOIL), to solve the optimization problem. Specifically, DPPOIL combines the proximal policy optimization algorithm \cite{schulman2017proximal} and generative adversarial imitation learning technique \cite{ho2016generative} to make it more suitable for the formulated optimization problem.
	
	\item Simulation results verify the effectiveness of DPPOIL in solving the TCRAMOP. Furthermore, the imitation learning is proven be efficient to improve the policy performance, enabling the DPPOIL to learn the best strategy compared with other baseline methods. 
\end{itemize}

\section{System Model and Problem Formulation}
\par
\begin{figure}[t]
	\centering
	\includegraphics[width=8cm]{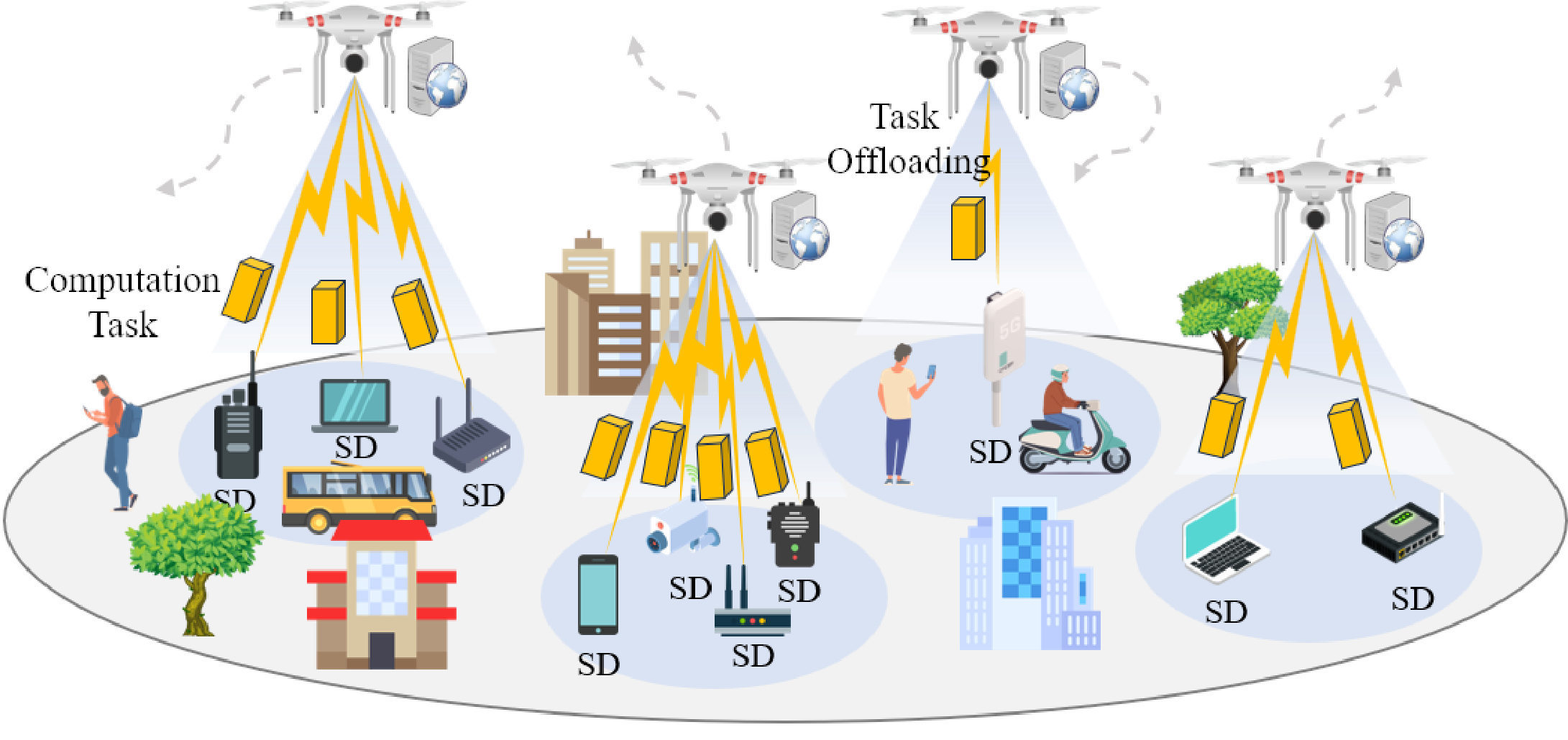}
	\caption{Multi-UAV-assisted MEC system.}
	\label{fig:system_model}
\end{figure}

 As shown in \figref{fig:system_model}, we consider a UAV-assisted MEC system which consists of multiple UAVs and a set of SDs. The SDs denoted by $\mathcal{M}\triangleq\{1,2,\dots,M\}$ are randomly distributed in a square area and can generate the computation tasks. Then, the UAVs are equipped with MEC servers and deployed to provide the computation service for SDs, which are represented as $\mathcal{N}\triangleq\{1,2,\dots,N\}$. Specifically, we assume that the UAV flies at a fixed altitude $H > 0$. When the UAV flies to a certain location, it will cover some SDs. Then, the UAV can simultaneously collect multiple tasks from the covered SDs and allocate a certain amount of computational resource to each task for computing these tasks. After completing the task computation, the UAV can flexibly adjust its position to effectively serve other devices. 

\par In addition, the system operates in the discrete time slot manner with $T$ equal time slots $\mathcal{T}=\{1, 2, \dots, T\}$, where each time slot has the identical time duration $\tau$. Then, the position of UAV $n$ at time slot $t$ can be denoted as $(x^{t}_{n}, y^{t}_{n}, H)$. As for SD, its position is fixed, and then its horizontal coordinate is expressed as $(x_{m}, y_{m})$.

\subsection{UAV Mobility Model}
\par In the work, we utilize the rotary-wing UAV due to its ability to hover and flexibly adjust its position in all directions. Furthermore, the UAV is assumed to fly at a fixed height $H \geq 0$. Then, the adjustment of position of UAV only requires to determine its horizontal flight direction and flight distance $[\theta^{t}_{n}, d^{t}_{n}]$. Therefore, the position of UAV at time slot $t$ can be expressed as
\begin{equation}\label{eq:UAV_position_x}
	x^{t}_{n} = x^{t-1}_{n} + d^{t-1}_{n}\cos\theta^{t-1}_{n}, 
\end{equation}
\begin{equation}\label{eq:UAV_position_y}
	y^{t}_{n} = y^{t-1}_{n} + d^{t-1}_{n}\sin\theta^{t-1}_{n}. 
\end{equation}
During the flight, the UAV flies at a constant speed $v^{t}_{n}=d^{t}_{n}/\tau$, which also has the maximum flight velocity constraint $v_{max}$. Then, the propulsion power consumption of UAV with respect to speed $v$ can be defined as \cite{zeng2019energy}
\begin{equation}\label{eq:energy_comsumption_2D}
	\begin{aligned}
		P(v) = & P_B\left(1 + \frac{3v^2}{v^2_{tip}}\right) + P_I\left(\sqrt{1+\frac{v^4}{4v^4_0}} - \frac{v^2}{2v^2_0}\right)^{1/2} \\& + \frac{1}{2}d_0 \rho sAv^3,
	\end{aligned}
\end{equation} 
where $P_{B}$ and $P_{I}$ are the blade profile power and the induced power, respectively. Moreover, $v_{tip}$ and $v_{0}$ separately denote the tip speed of rotor blade and the mean rotor induced velocity. Additionally, $d_{0}$, $\rho$, $s$, and $A$ are the fuselage drag ratio, air density, rotor solidity and rotor disc area, respectively.

\subsection{Communication Model}
\par The UAV has a communication coverage with radius $R$, and then some closest SDs to the UAV within the coverage area can establish the uplinks with UAV so as to transmit their computation tasks to UAV. Furthermore, the UAV adopts the orthogonal frequency division multiple access (OFDMA) technique, which can enable multiple SDs to transmit simultaneously. However, due to limited resource, a single UAV can connect to up to $S$ devices at once. Then, the transmission rate between the SD $m$ and the UAV $n$ at time slot $t$ can be expressed as
\begin{equation}\label{eq:SD_transmission_rate}
	R^{t}_{m, n} = W\log_2\left(1 + \frac{P_{t}g^{t}_{c, mn}}{\sigma^2}\right),
\end{equation}
where $W$ is the bandwidth allocated to each SD, $P_{t}$ is the transmit power of SD, $\sigma^2$ is the noise power, and $g^{t}_{c, mn}$ is the channel power gain between the SD $m$ and the UAV $n$. 

\par Benefiting from the high flight altitude of UAV, the SD has the large probabilty to obtain the LoS link. Hence, the channel power gain $g^{t}_{c, mn}$ is modeled by the probabilistic LoS, which is expressed as 
\begin{equation}\label{eq:channel_power_gain}
	g^{t}_{c, mn} = K^{-1}_{o}{d^{t}_{m,n}}^{-\alpha}[P^{t}_{\mathrm{LoS}}\mu_{\mathrm{LoS}} + P^{t}_{\mathrm{NLoS}}\mu_{\mathrm{NLoS}}]^{-1},
\end{equation}
where $d^{t}_{m,n}$ and $\alpha$ denote the distance between the SD and the UAV as well as the path loss exponent, respectively. $K_{o}=(\frac{4 \pi f_{c}}{c})^{2}$, where $f_{c}$ is the carrier frequency, and $c$ is the speed of light. $P^{t}_{\mathrm{LoS}}$ and  $P^{t}_{\mathrm{NLoS}}=1-P^{t}_{\mathrm{LoS}}$ separately represent the LoS probabilty and the non-LoS (NLoS) probability, and then $\mu_{\mathrm{LoS}}$ and $\mu_{\mathrm{NLoS}}$ are different attenuation factors for the LoS and NLoS links. 

\par For the LoS probability $P^{t}_{\mathrm{LoS}}$, except that it is influenced by the transmission height, it is also associated with the propagation environment as well as the density and height of buildings. Thus, the LoS probability $P^{t}_{\mathrm{LoS}}$ is modeled as \cite{al2014optimal}
\begin{equation}\label{eq:LoS_probability}
	P^{t}_{\mathrm{LoS}} = \frac{1}{1+c_{1}\operatorname{exp}(-c_{2}(\theta^{t}_{m,n}-c_{1}))},
\end{equation} 
where $c_{1}$ and $c_{2}$ are the parameters related to the environment. $\theta^{t}_{m,n} = \frac{180}{\pi} \times \sin^{-1}(\frac{H}{d^{t}_{m,n}})$ is the elevation angle between the SD and the UAV.

\subsection{Computation Model}
\par We assume that each SD can generate a computation task at each time slot, which is defined as $<D^{t}_{m}, C^{t}_{m}, \tau^{t}_{m}>$, where $D^{t}_{m}$ is the task size,  $C^{t}_{m}$ is the computation intensity of task (cycles/bit), and $\tau^{t}_{m}$ is the deadline of the task. When a UAV is connected to several SDs at time slot $t$, these SDs can upload their tasks to UAV, where the set of served SDs is denoted as $\mathcal{M}^{t}_{n}$ and $|\mathcal{M}^{t}_{n}| \leq S$. Then, the delay for the SD to transmit the task to the UAV is given as
\begin{equation}\label{eq:transmission_delay}
	T^{trans}_{m, n, t} = \frac{D^{t}_{m}}{R^{t}_{m, n}}.
\end{equation} 
After uploading tasks, the UAV will allocate a certain amount of computation resource to each task for computation. Hence, the delay in processing task by UAV can be denoted as 
\begin{equation}\label{eq:computation_delay}
	T^{comp}_{n, m, t} = \frac{D^{t}_{m}C^{t}_{m}}{f^{t}_{n, m}},
\end{equation} 
where $f^{t}_{n, m}$ is the computing resource allocated by UAV $n$ to the SD $m$. Then, the total delay for UAV $n$ to offload the task of SD $m$ is expressed as $T^{off}_{n, m, t} = \frac{D^{t}_{m}}{R^{t}_{m, n}} + \frac{D^{t}_{m}C^{t}_{m}}{f^{t}_{n, m}}$.
In sum, at time slot $t$, the total delay comsumed by UAV $n$ can be represented as $T^{total}_{n, t} = \sum_{m\in \mathcal{M}^{t}_{n}} T^{off}_{n, m, t}$.

\par For the computation energy consumption of UAV, the energy consumed by the UAV $n$ to compute the task of SD $m$ at time slot $t$ can be denoted as \cite{lyu2018energy}
\begin{equation}\label{eq:computation_energy}
	E^{comp}_{n, m, t} = \kappa_{m}(f^{t}_{n, m})^{2}D^{t}_{m}C^{t}_{m},
\end{equation}
where $\kappa_{m} \geq 0$ is the effective capacitance of CPU of UAV $n$ that is related to the CPU architecture \cite{pan2021cost}. Accordingly, the total computation energy consumption of UAV $n$ at time slot $t$ is expressed as $E^{comp}_{n, t} = \sum_{m\in \mathcal{M}^{t}_{n}} E^{comp}_{n, m, t}$.

\subsection{Problem Formulation}
\par In the work, we aim to improve the MEC performance of UAVs while reducing the UAV energy consumption. In addition, although the OFDMA technique eliminates the intra-cell communication interference among SDs served by the same UAV, the inter-cell interference from SDs served by other UAVs may exist if all UAVs move throughout the whole region. Hence, to prevent this interference, we divide the whole area into several sub-areas, and then each UAV is assigned to a sub-area. 

\par To improve the performance of UAV MEC service, we consider increasing the offloading number of UAVs and decreasing the offloading delay of tasks. Furthermore, to prevent a single UAV from repeatedly serving the same SD, according to \cite{li2023computing}, we define a fairness indicator which is expressed as $f_{m, t} = 1 - \lambda \frac{b_{m, t}}{T}$,
where $b_{m, t}$ denotes the offloading number of SD $m$ from time slot 1 to time slot $t$, and $\lambda$ is a scaling factor. In sum, to increase the offloading number and guarantee the fairness among SDs, we optimize the fairness index about offloading number of all UAVs at time slot $t$, which is defined as $F_{t} = \sum_{n=1}^{N}\sum_{m\in \mathcal{M}^{t}_{n}} f_{m, t}$.
Then, the offloading delay required to be optimized is expressed as 
$T^{Delay}_{t} = \sum_{n=1}^{N}\sum_{m\in \mathcal{M}^{t}_{n}} T^{off}_{n, m, t}$.
For optimization to energy consumption, the mobility and computation energy consumption need to be considered. Then, the energy consumption of all UAVs at time slot $t$ is denoted as $E_{t} = \sum_{n=1}^{N}P(v^{t}_{n})\tau + E^{comp}_{n, t}$, where $v^{t}_{n}$ is the flight speed of UAV $n$ at current time slot.

\par Thus, the considered three optimization objectives are mathematically denoted as $F_{\mathrm{total}}=\sum^{T}_{t=1}F_{t}$,  $T_{\mathrm{total}}=\sum^{T}_{t=1}T^{Delay}_{t}$ and $E_{\mathrm{total}}=\sum^{T}_{t=1}E_{t}$, which are the total fairness index about offloading number of UAVs, the total offloading delay and the total energy consumption of UAVs, respectively. Then, the three objectives are determined by the UAV trajectories and the allocated computation resource of served SDs, which are characterized by the sequences of horizontal positions $\boldsymbol{\mathrm{X}}=\{x^{t}_{n}, \forall n \in \mathcal{N}, \forall t \in \mathcal{T}\}$,  $\boldsymbol{\mathrm{Y}}=\{y^{t}_{n}, \forall n \in \mathcal{N}, \forall t \in \mathcal{T}\}$, and the sequences about computation resource allocated to each served SD $\boldsymbol{\mathrm{F}}=\{f^{t}_{n, m}, \forall n \in \mathcal{N}, \forall t \in \mathcal{T}, \forall m \in \mathcal{M}^{t}_{n} \}$. In summary, the TCRAMOP is formulated as follows:
\begin{subequations}
	\begin{align}
		&\mathop{\operatorname{max}}\limits_{\boldsymbol{\mathrm{X}}, \boldsymbol{\mathrm{Y}}, \boldsymbol{\mathrm{F}}} & &  f=\{F_{\mathrm{total}}, -T_{\mathrm{total}}, -E_{\mathrm{total}}\}\label{optimization_problem_1}  \\
		&\quad~~\textrm{s.t.} & & 0 \leq x^{t}_{n} \leq L, \forall n\in\mathcal{N},  \forall t\in\mathcal{T}  \label{optimization_problem_2},\\
		&&&0 \leq y^{t}_{n} \leq L, \forall n\in\mathcal{N},  \forall t\in\mathcal{T}  \label{optimization_problem_3},\\
		&&&0 \leq \theta^{t}_{n} \leq 2\pi, \forall n\in\mathcal{N},  \forall t\in\mathcal{T}  \label{optimization_problem_4},\\
		&&&0 \leq d^{t}_{n} \leq d_{max}, \forall n\in\mathcal{N},  \forall t\in\mathcal{T} \label{optimization_problem_5},\\
		&&&0 \leq f^{t}_{n, m} \leq f^{max}_{n}, \forall n\in\mathcal{N}, \forall t\in\mathcal{T} \label{optimization_problem_6},\\
		&&&\sum\nolimits_{m\in \mathcal{M}^{t}_{n}} f^{t}_{n, m} \leq f^{max}_{n}, \forall n\in\mathcal{N},  \forall t\in\mathcal{T} \label{optimization_problem_7}, \\
		&&& T^{off}_{n, m, t} \leq \tau^{t}_{m}, \forall n\in\mathcal{N}, \forall m\in \mathcal{M}^{t}_{n},  \forall t\in\mathcal{T} \label{optimization_problem_8}.
	\end{align} 
	\label{optimization_problem}
\end{subequations}
The formulated TCRAMOP is a non-convex multi-objective optimization problem with conflicting objectives. Additionally, the system is dynamic and the UAVs require to make continuous decisions. To tackle these challenges, we adopt the DRL approach to address the problem.

\section{Proposed DRL approach}
\par Consider that multiple UAVs independently carry out the tasks, we adopt the DRL approach to solve our optimization problem without considering the multi-agent setting. In addition to using the PPO algorithm \cite{schulman2017proximal}, we also combine the generative adversarial imitation learning \cite{ho2016generative} technique to improve the algorithm performance, and then propose the DPPOIL algorithm to solve the TCRAMOP.

\subsection{Markov Game for TCRAMOP}
\par To use DRL in TCRAMOP, we formulate a multi-agent Markov Game with the set of agents $\mathcal{N} = \{1, \ldots, N\}$, observation space $\mathcal{O} = \{\mathcal{O}_{1}, \ldots, \mathcal{O}_{N}\}$, action space $\mathcal{A} = \{\mathcal{A}_{1}, \ldots, \mathcal{A}_{N}\}$ and reward space $\mathcal{R} = \{\mathcal{R}_{1}, \ldots, \mathcal{R}_{N}\}$. Then, the detailed constructions of above components are explained as follows.

\subsubsection{Observation Space}
\par At each time slot, each agent takes measure according to its obtained observation. Hence, the observation obtained by agent $n$ at time slot $t$ is expressed as
\begin{equation}\label{eq:observation_space}
	\begin{aligned}
		o^{t}_{n} = \{x^{t}_{n}, y^{t}_{n}, z^{t}_{n}, \{\theta^{t}_{m, n}, d^{t}_{m,n}, &b_{m,t}, f^{min}_{m, t}\}_{m \in \mathcal{M}^{S, t}_{n}},  \\
		&\{\theta^{t}_{k, n}, d^{t}_{k,n}\}_{k \in \mathcal{N}^{t}_{n, q}}\},
	\end{aligned}
\end{equation}
where $\mathcal{M}^{S, t}_{n}$ denotes the set of SDs that can be observed by UAV $n$. $\theta^{t}_{m, n}$ and $d^{t}_{m,n}$ are separately the horizontal direction and the horizontal distance between the SD $m$ and UAV $n$. Then, $b_{m,t}$ denotes the cumulative offloading times of SD $m$ from time slot 1 to time slot $t$, and $f^{min}_{m, t}$ is the minimum computing resource required for the current task of SD $m$. In addition, $\mathcal{N}^{t}_{n, q}$ represents the set of $q$ nearest UAVs to UAV $n$. Then, $\theta^{t}_{k, n}$ and $d^{t}_{k,n}$ denote the horizontal direction and the horizontal distance between the UAV $k$ and UAV $n$, respectively.

\subsubsection{Action Space}
\par In the system, when the state of environment changes, the UAV will adjust its position and allocate its computing resource to linked SDs. Therefore, the action of agent $n$ at time slot $t$ is denoted as
\begin{equation}\label{eq:action_space}
	\begin{aligned}
		a^{t}_{n} = \{\theta^{t}_{n}, d^{t}_{n}, f^{t}_{n, 1},\dots, f^{t}_{n, m},\dots,f^{t}_{n, S} \},
	\end{aligned}
\end{equation}
where $S$ represents the maximum number of SDs served by UAV.

\subsubsection{Reward Function}
\par Since the UAV-assisted MEC system intends to maximize the total fairness index about offloading number of UAVs and minimize the total offloading delay as well as the total energy consumption of UAVs, the corresponding reward of agent $n$ at time slot $t$ is defined as:
\begin{equation} 
	\label{eq: the_reward_function}
	\begin{aligned}
		r^{E, t}_{n} &=  \omega_{1} \underbrace{\sum_{m\in \widetilde{\mathcal{M}}^{t}_{n}} f_{m, t} \log_2\left(1 + \frac{P_{t}g^{t}_{c, mn}}{\sigma^2}\right)}_{r^{\mathrm{Offload}, t}_{n}} + \omega_{2}r^{\mathrm{Num}, t}_{n} \\ &+ \omega_{3}\underbrace{\sum_{m\in \widetilde{\mathcal{M}}^{t}_{n}} f^{t}_{n, m}}_{r^{\mathrm{Res}, t}_{n}} - \omega_{4}\underbrace{P(v^{t}_{n})\tau}_{r^{\mathrm{Move}, t}_{n}} - \omega_{5}\underbrace{\sum_{m\in \widetilde{\mathcal{M}}^{t}_{n}}  E^{comp}_{n, m, t}}_{r^{\mathrm{Comp}, t}_{n}} \\ & + \omega_{6}\underbrace{\sum_{k\in \mathcal{N}, k\neq n}d^{t}_{k,n}}_{r^{\mathrm{U2U}, t}_{n}},
	\end{aligned}
\end{equation}
where $\omega_{1}$, $\omega_{2}$, $\omega_{3}$, $\omega_{4}$, $\omega_{5}$ and $\omega_{6}$ are separately the weight coefficient of each reward, and $\widetilde{\mathcal{M}}^{t}_{n}$ denotes the set of SDs connected by UAV $n$ at time slot $t$. Furthermore, $r^{\mathrm{Offload}, t}_{n}$ represents the reward for the fairness index and the transmission rate of SDs connected to UAV $n$. Then, $r^{\mathrm{Num}, t}_{n}$ is the reward for the number of SDs covered by UAV $n$. Moreover, $r^{\mathrm{Res}, t}_{n}$ denotes the reward for the computing resource allocated to connected SDs. In addition, $r^{\mathrm{Move}, t}_{n}$ and $r^{\mathrm{Comp}, t}_{n}$ are penalties for moving energy consumption and computation energy consumption of UAV $n$, respectively. Then, $r^{\mathrm{U2U}, t}_{n}$ is the reward for the distances between the UAV $n$ and other UAVs.

\subsection{Basic PPO}
\par The PPO algorithm is an on-policy algorithm from policy gradient method, which shows the robustness and the efficiency on a variety of tasks \cite{schulman2017proximal}. Specifically,  there is an actor network $\pi_{\theta}(a^{t}_{n}|o^{t}_{n})$ and a critic network $V_{\phi}(o^{t}_{n})$ in PPO. Then, the actor network $\pi_{\theta}(a^{t}_{n}|o^{t}_{n})$ is responsible for selecting actions $a^{t}_{n}$ based on the current observation $o^{t}_{n}$. Furthermore, the update of actor network is to minimize a surrogate loss function, which is expressed as 
\begin{equation}
	\label{eq: actor update}
	\begin{aligned}
		L_{A}(\theta)= \frac{1}{B}\sum^{B}_{b=1}&\operatorname{min}\Big[\frac{\pi_{\theta}(a_{b} | o_{b})}{\pi_{\theta^{\text{old}}}(a_{b} | o_{b})}A_{b}, \\&\operatorname{clip}(\frac{\pi_{\theta}(a_{b} | o_{b})}{\pi_{\theta^{\text{old}}}(a_{b} | o_{b})}, 1-\epsilon, 1+\epsilon)A_{b}\Big],
	\end{aligned}
\end{equation}
where $B$ denotes the sampled batch size and $\frac{\pi_{\theta}(a_{b} | o_{b})}{\pi_{\theta^{\text{old}}}(a_{b} | o_{b})}$ is the probability ratio between the current policy $\pi_{\theta}$ and the old policy $\pi_{\theta^{\text{old}}}$. Then, the $\operatorname{clip}(\cdot)$ operator constrains the update of policy to a certain range $(1-\epsilon, 1+\epsilon)$, where $\epsilon$ is a hyperparameter to control the difference between the current policy $\pi_{\theta}$ and the old one $\pi_{\theta^{\text{old}}}$. Moreover, $A_{b}$ is the computed advantage function.

\par The critic network $V_{\phi}(o^{t}_{n})$ is used for estimating the state-value of observation $o^{t}_{n}$. Moreover, its update adopts the conventional mean squared error (MSE) loss function with respect to the predicted value $V_{\phi}(o_{n}^{t})$ and the discounted return $R_{n}^{t}$ about reward $r^{t}_{n}$, which is given by
\begin{equation}
	\label{eq: critic update}
	L_V(\phi) = \frac{1}{2B}\sum^{B}_{b=1}\left( V_{\phi}(o_{b}) - R_{b}\right)^2.
\end{equation}

\begin{figure*}[t]
	\centering
	\includegraphics[width=1\linewidth]{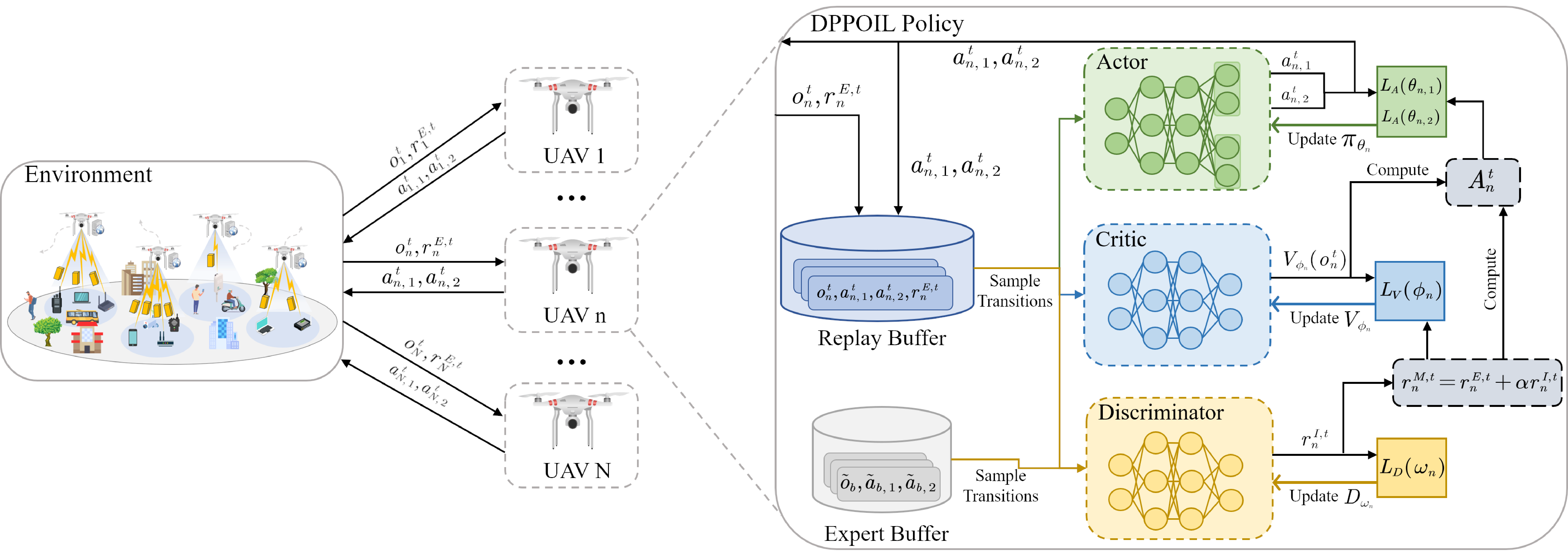}
	\caption{Flowchart of the proposed DPPOIL algorithm.}
	\label{fig:algorithm_flowchart}
\end{figure*}

\subsection{DPPOIL Framework}
\par DPPOIL algorithm combines the PPO algorithm and the generative adversarial imitation learning technique. Hence, each agent has an actor network $\pi_{\theta_{n}}(a^{t}_{n}|o^{t}_{n})$, a critic network $V_{\phi_{n}}(o^{t}_{n})$, and a discriminator network $D_{\omega_{n}}(o^{t}_{n}, a^{t}_{n})$. Furthermore, the UAV needs to allocate the computing resource to SDs and the sum of allocated computing resource does not exceed the on-board computational resource of UAV. Then, the actor network has two policies, where the first policy $\pi_{\theta_{n, 1}}(a^{t}_{n, 1}|o^{t}_{n})$ uses the Gaussian distribution to sample the UAV flight actions and the second policy $\pi_{\theta_{n, 2}}(a^{t}_{n, 2}|o^{t}_{n})$ adopts the Dirichlet distribution to sample the actions about the computing resource allocated to SDs, whose action elements sum to 1. In addition, the discriminator network is used to judge whether the current generated strategy comes from the expert policy or the current policy of agent. Moreover, the discriminator network can produce the intrinsic reward $r^{I,t}_{n}=\log(D_{\omega_{n}}(o^{t}_{n}, a^{t}_{n, 1}, a^{t}_{n, 2}))$ according to the current practical performance of agent, which will participate in the algorithm training with the extrinsic reward $r^{E,t}_{n}$ to improve the policy performance.

\par As for training algorithm, a normal replay buffer $\mathcal{B}=\{\textbf{o}, \textbf{a}_{\textbf{1}}, \textbf{a}_{\textbf{2}}, \textbf{r}^{\textbf{E}}\}$ and an expert buffer $\widetilde{\mathcal{B}}=\{\tilde{\textbf{o}}, \tilde{\textbf{a}}_{\textbf{1}}, \tilde{\textbf{a}}_{\textbf{2}}\}$ are provided for training these networks, where the expert buffer $\widetilde{\mathcal{B}}$ is specialized for training discriminator network. Specifically, 
the discriminator network is the first to start. Then, the discriminator network is trained for distinguishing whether the current policy is the expert policy or the agent policy. Hence, the optimization to discriminator network requires to measure the binary cross entropy about the agent policy and the expert policy, and then the loss function is expressed as
\begin{equation}\label{eq:exploration_module_objective}
	\begin{aligned}
		L_{D}(\omega_{n})=&-\frac{1}{B}\sum^{B}_{b=1}\log(D_{\omega_{n}}(o_{b}, a_{b, 1}, a_{b, 2})) \\ &-\frac{1}{\widetilde{B}}\sum^{\widetilde{B}}_{b=1}(1-\log(D_{\omega_{n}}(\tilde{o}_{b}, \tilde{a}_{b, 1}, \tilde{a}_{b, 2}))).
	\end{aligned}
\end{equation} 
After training the discriminator network, the discriminator network can generate the intrinsic reward $r^{I,t}_{n}=\log(D_{\omega_{n}}(o^{t}_{n}, a^{t}_{n, 1}, a^{t}_{n, 2}))$, and it will be combined with the extrinsic reward $r^{E,t}_{n}$ to form a mixed reward $r^{M,t}_{n}=r^{E,t}_{n} + \alpha r^{I,t}_{n}$ so as to participate in the training of actor network and critic network, where $\alpha$ is a scaling coefficient. Besides, the training of actor network and critic network adopts the regular training patterns of PPO. Furthermore, since the actor network has two action policies, the two policies separately have a loss function (i.e., $L_{A}(\theta_{n, 1})$ and $L_{A}(\theta_{n, 2})$). Then, the overall structure and flow of DPPOIL framework are visually illustrated in \figref{fig:algorithm_flowchart}.

\section{Simulation Results}
\par In the section, we evaluate the performance of proposed DPPOIL in solving the TCRAMOP.

\par In the system setting, we consider a square area with 1000 m $\times$ 1000 m, where the 100 SDs are distributed in the area. Then, consider the interference among UAVs, the square area is evenly divided into four sub-square regions, and four UAVs separately regulate a sub-region. The task execution duration is set to 150 s and is equally divided into $T=30$ time slots with length of $\tau=5$ s. The settings about other key parameters are summarized in \tabref{table:parameter settings}.

\par In DRL setting, the proposed DPPOIL runs 60000 episodes in total, where each episode has 30 time slots. Then, in reward function, the weight coefficients of
reward components $\omega_{1}$, $\omega_{2}$, $\omega_{3}$, $\omega_{4}$, $\omega_{5}$, and $\omega_{6}$ are set to 100, 5, 20, 20, 10, and 1, respectively. In addition, all networks are made up of two hidden layers, and the dim of each hiddern layer is 64. Moreover, the optimizer uses the Adam optimizer, and the trainings of all networks adopt 0.0005 learning rate.

\begin{table}[t]
	\renewcommand\arraystretch{1.0}
	\setlength\tabcolsep{0pt}
	\centering 
	\caption{Parameter settings}
	\label{table:parameter settings}
	
	\begin{tabular}{cc}\toprule[1.5pt]
		\textbf{Parameter} & \textbf{Value} \\ \toprule[1pt]
		UAV mass $(m_{\mathrm{UAV}})$ & 2 kg \\
		UAV flight altitude $(H)$& 120 m \\
		Maximum flight distance of UAV $(d_{max})$ & 150 m \\
		Computing resource of UAV MEC server $(f^{max}_{n})$ & 20 Ghz \\
		Transmit power of SD $(P_{t})$ & 0.1 W \\
		Task size & [1, 5] Mb \\
		Computation intensity of task & [500, 1500] cycles/bit \\
		Deadline of task & [1, 5] s \\
		\bottomrule[1.5pt]
	\end{tabular}
\end{table}

\par To thoroughly investigate the performance of DPPOIL, we implement five baseline methods for comparison. These approaches are separately UAVs with random strategy (Random), deep deterministic policy gradient (DDPG) algorithm, multi-agent deep deterministic policy gradient (MADDPG) algorithm, multi-agent proximal policy optimization (MAPPO) algorithm and the
conventional PPO algorithm. Furthermore, we have continuously tuned these DRL algorithms to guarantee their convergence.

\subsection{Convergence Analysis}
\par
\begin{figure*}
	\centering
	\subfloat[]{
		\includegraphics[width=0.3\linewidth]{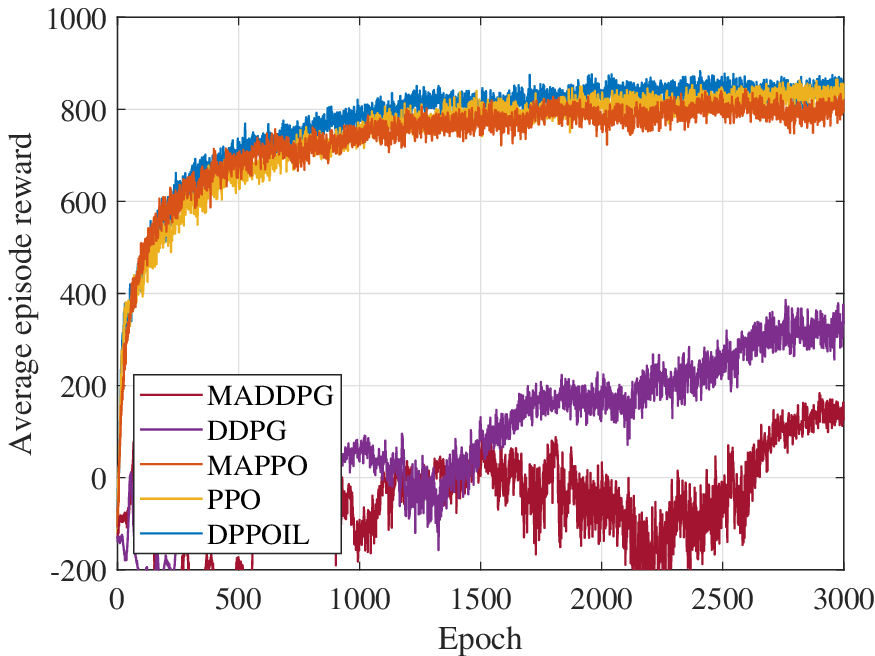}\label{fig:convergence_performance}}
	\subfloat[]{
		\includegraphics[width=0.38\linewidth]{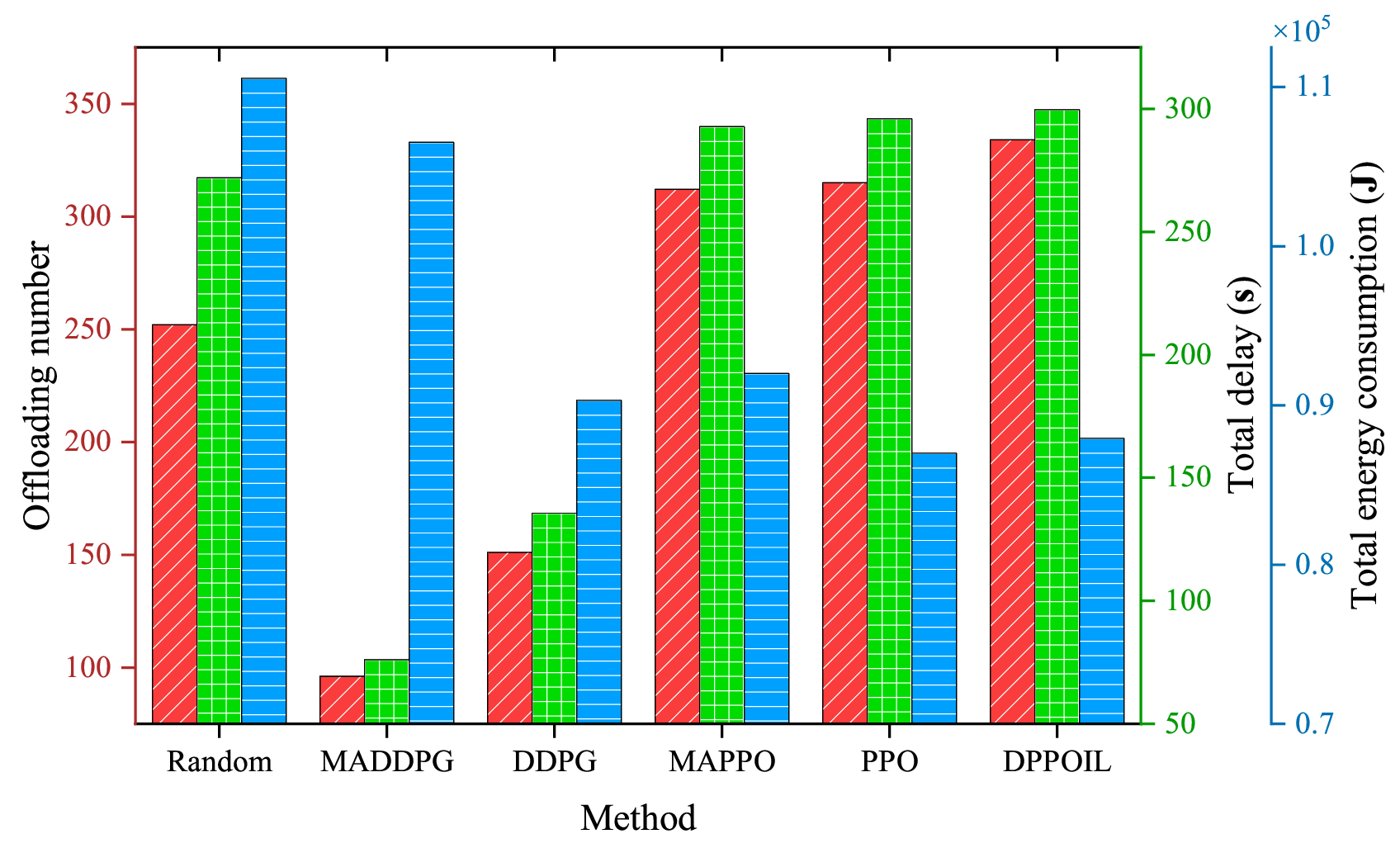}\label{fig:optimization_results}}
	\subfloat[]{
		\includegraphics[width=0.3\linewidth]{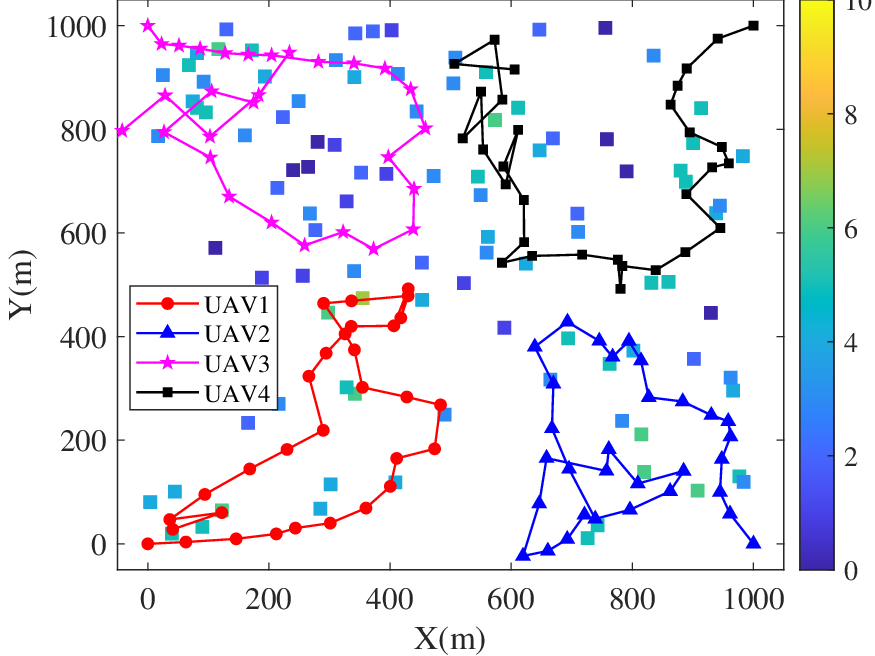}\label{fig:practical_performance}} 
	
	\caption{Simulation results. (a) Convergence performance under various approaches. (b) Optimization results of six methods. (c) Flight trajectories of UAVs optimized by DPPOIL.}
	\label{fig:results}
\end{figure*}

 We evaluate the convergence performance of DPPOIL and other methods in the section. Fig. \ref{fig:results}(a) shows the convergence of all DRL approaches with respect to the extrinsic reward. It can be seen that DPPOIL, PPO and MAPPO achieve the outstanding performance compared with DDPG and MADDPG. Specifically, the three algorithms have the same convergence speed, starting to converge at about 1500 epoch. Then, DPPOIL has the best performance, whose ultimate convergence level is around 850 and learning curve is always higher than PPO. As for MAPPO, it does not perform as well as the other two algorithms, whose performance is not better than PPO. Meanwhile, the performance of MADDPG that is a multi-agent variant of DDPG is also worse than single-agent DDPG. 

%

\subsection{Performance Comparison}
\par We also test the practical performance of these algorithms as UAVs perform an identical MEC mission. Fig. \ref{fig:results}(b) shows the optimization results of the six methods. It can be observed that the proposed DPPOIL outperforms all other methods in the objective of offloading the most tasks, which means that DPPOIL learns the better computation resource allocation strategy and exploration strategy due to the enhancement of imitation learning. Moreover, benefiting from better resource allocation strategy, although the total delay of DPPOIL is larger than PPO, the corresponding average delay is lower than PPO. Then, PPO consume the less energy than DPPOIL, while it offloads the fewer tasks. In addition, the multi-agent algorithms (i.e., MAPPO, MADDPG) has the poorer performance in executing the MEC task compared with their corresponding single-agent algorithms (i.e., PPO, DDPG). Furthermore, due to the limitations of algorithms, DDPG and MADDPG have the markedly bad performance. Besides, Fig. \ref{fig:results}(c) displays the flight trajectories of UAVs optimized by DPPOIL during the MEC task, where the colored squares represent the SDs and various colors denote the offloading times of SDs. It can be seen that all UAVs can rationally provide the MEC services, which do not fly into the areas regulated by other UAVs and ensure the fairness among SDs.

%
%
%
%

\section{Conclusion}
\par In this paper, a UAV-assisted MEC system is studied, where multiple UAVs jointly perform the edge computing tasks and a single UAV can simultaneously offer the computing support for multiple SDs. Then, we formulated a TCRAMOP to simultaneously maximize the offloading number of UAVs and minimize total offloading delay and total energy consumption of UAVs. Nevertheless, consider that the UAVs need to execute continous decisions and the system is dynamic, we proposed a DPPOIL that combines the conventional PPO algorithm and generative adversarial imitation learning technique to solve the problem. Simulation results shown that the proposed DPPOIL achieves the outstanding performance on conducting the MEC task. Moreover, it is verified that the imitation learning improves the policy performance, and then the strategy of DPPOIL is better than other baseline methods.

\bibliographystyle{IEEEtran}
\bibliography{MEC}

\end{document}